# A Kinetic Study of the N($^2$D) + C$_2$H$_4$ Reaction at Low Temperature

Kevin M. Hickson,[a,b] Cédric Bray,[a,b] Jean-Christophe Loison,[a,b] and Michel Dobrijevic[c]

[a]*Université de Bordeaux, Institut des Sciences Moléculaires, UMR 5255, F-33400 Talence, France*

[b]*CNRS, Institut des Sciences Moléculaires ,UMR 5255, F-33400 Talence, France*

[c]*Laboratoire d'Astrophysique de Bordeaux, Université de Bordeaux, CNRS, B18N, allée Geoffroy Saint-Hilaire, F-33615 Pessac, France*

**Abstract**

Electronically excited nitrogen atoms N($^2$D) are important species in the photochemistry of N$_2$ based planetary atmospheres such as Titan. Despite this, few N($^2$D) reactions have been studied over the appropriate low temperature range. During the present work, rate constants were measured for the N($^2$D) + ethene (C$_2$H$_4$) reaction using a supersonic flow reactor at temperatures between 50 K and 296 K. Here, a chemical reaction was used to generate N($^2$D) atoms, which were detected directly by laser induced fluorescence in the vacuum ultraviolet wavelength region. The measured rate constants displayed very little variation as a function of temperature, with substantially larger values than those obtained in previous work. Indeed, considering an average temperature of 170 K for the atmosphere of Titan leads to a rate constant that is almost seven times larger than the currently recommended value. In parallel, electronic structure calculations were performed to provide insight into the reactive process. While earlier theoretical work at a lower level predicted the presence of a barrier for the N($^2$D) + C$_2$H$_4$ reaction, the present calculations demonstrate that two of the five doublet potential energy surfaces correlating with reagents are likely to be attractive, presenting no barriers for the perpendicular approach of the N atom to the C=C bond of ethene. The measured rate constants and new product channels taken from recent dynamical investigations of this process are included in a 1D coupled ion-neutral model of Titan's atmosphere. These simulations indicate that the modeled abundances of numerous nitrogen bearing compounds are noticeably affected by these changes.

# 1 Introduction

As the fifth most abundant element in the Universe, nitrogen and its compounds make an important contribution to the chemistry of a wide range of different environments such as the interstellar medium and planetary atmospheres. In its ground electronic state configuration, $^4S_{3/2}$, atomic nitrogen is unreactive with most stable molecules at low and ambient temperature although it does display a significant reactivity towards radical species.[1-8]

Unlike N($^4$S) atoms, experimental studies have shown that atomic nitrogen in its first excited electronic state, N($^2$D) reacts much more rapidly with closed shell molecules.[9, 10] Moreover, as its electronic state $^2D_{3/2}$ and $^2D_{5/2}$ configurations are characterized by long radiative lifetimes of 13.6 and 36.7 hours respectively,[11] these atoms are expected to play an important role in the chemistry of planetary atmospheres containing molecular nitrogen such as the Earth, Pluto and Saturn's moon Titan. In these environments, solar radiation is absorbed by atmospheric $N_2$ in the vacuum ultraviolet (VUV) range, leading to N($^2$D) + N($^4$S) formation below 102 nm.[12] Higher excited states such as N($^2$P) are also produced at wavelengths shorter than 93 nm, although these atoms are significantly less reactive than N($^2$D) and are mostly removed by quenching.[9, 10] In Titan's atmosphere, which is characterized by a large percentage of molecular nitrogen (95 %), the reactions of N($^2$D) atoms with hydrocarbon species such as methane (present at the 2-5 % level) and other larger saturated and unsaturated hydrocarbons present at trace levels are thought to drive much of the chemistry above 800 km. Indeed, N($^2$D) atoms formed at these altitudes are relatively unaffected by non-reactive losses due to their long radiative lifetime and the slow electronic quenching of N($^2$D) by $N_2$ at low temperature.[13, 14] Previous work on the kinetics[15-24] and dynamics[25-30] of N($^2$D) atom reactions with saturated and unsaturated hydrocarbons indicate that these processes could lead to significant quantities of nitrogen bearing organic molecules. Indeed, several reactions involving N($^2$D) atoms have been identified as potentially key processes in earlier photochemical modeling studies of Titan's atmosphere.[31-33] Despite this, there are few measurements of the rate constants for N($^2$D) reactions at relevant temperatures (70 - 170 K), particularly with the most abundant species such as $CH_4$, $C_2H_6$, $C_2H_4$, $C_2H_2$ and HCN.

Nunez-Reyez et al.[22] examined the reactivity of N($^2$D) with saturated hydrocarbon molecules over the 75–296 K temperature range. Measured rate constants for the N($^2$D) + $CH_4$ reaction were seen to agree with earlier work over the 223-293 K range,[15-17, 19] thereby validating the values recommended by Herron[9] and Dutuit et al.[10] In contrast, the recommended rate constants for the N($^2$D) + $C_2H_6$ and $C_3H_8$ reactions, based on room temperature experiments, were found to be significantly larger than the rate constants measured by Nunez-Reyes et al.[22] at temperatures relevant to Titan's atmosphere. These authors showed that both of these reactions are less efficient as the temperature falls, becoming negligibly important to Titan's photochemistry. In later work, Nunez-Reyes et al.[23] also studied the kinetics of the $N(^2D)$ + $C_2H_2$ reaction. The measured rate constants were seen to remain large and essentially temperature independent over the 50-296 K range, while new ab-initio calculations confirmed the barrierless nature of the reaction over the $^2B_1$ potential energy surface (PES). These results were in contrast with earlier work by Takayanagi et al.[21] over the 223-293 K range, who reported a decrease in the reactivity of this process at low temperature. Extrapolation of the Takayanagi et al.[21] results to temperatures relevant to Titan's atmosphere (170 K) predicted a rate constant more than 3 times lower than the Nunez-Reyes et al.[23] result. When the Nunez-Reyes et al.[23] rate constants are included in a photochemical model of Titan's atmosphere instead of the recommended values[9] (based on the Takayanagi et al.[21] values), large increases in the mole fractions of related species such as HCCN (product of the $N(^2D)$ + $C_2H_2$ reaction) and CCN (product of the H + HCCN reaction) are predicted.

The studies of Nunez-Reyes et al.[22, 23] show a clear need for additional measurements of $N(^2D)$ reactions with other abundant unsaturated hydrocarbons species such as $C_2H_4$. The kinetics of the $N(^2D)$ + $C_2H_4$ reaction have been previously studied at room temperature[15, 16, 20] and from 230 - 292 K by Sato et al.,[24] who obtained rate constants that decrease to low temperature. The temperature dependence of the Sato et al.[24] results were rationalized in the context of quantum-chemical calculations[24, 34] that predicted the presence of a barriers over the PES leading from reagents to products. Several different product channels are possible for the $N(^2D)$ + $C_2H_4$ reaction as determined by crossed molecular beam experiments[26, 27, 30] and theoretical calculations.[27, 30, 34, 35] Yields as recommended by Dutuit et al.[10] include the major products $CH_2NCH$ + H (67 %), $c$-$CH_2(N)CH$ + H (23 %) and $CH_2CNH$ + H (5 %) with other minor channels making up the remaining 5 %. Interestingly, the most recent quantum chemical calculations by Balucani et al.[27] at the CCSD(T)/aug-cc-pVTZ level seem to indicate that this process might not be characterized by a barrier in the entrance channel in contrast to earlier studies by Takayanagi et al.,[34] Sato et al.[24] and Lee et al.[30]

Here we present an experimental kinetics study of the the $N(^2D)$ + $C_2H_4$ reaction using a supersonic flow reactor over the 50-296 K temperature range. In common with our earlier studies of $N(^2D)$ reactions, these atoms were produced by a chemical reaction and detected by laser induced fluorescence (LIF). Electronic structure calculations were performed to rationalize the measured rate constants and the presence, or absence, of a barrier in the entrance channel. Finally, the new rate constants for this reaction were introduced into an updated photochemical model of Titan's atmosphere to test its effect on species abundances. The chemical network used was reviewed as the products of the N($^2$D) + C$_2$H$_4$ reaction have been studied recently[27] showing that the most abundant products are not those currently considered in most models. The experimental and theoretical methods are presented in sections 2 and 3, followed by a discussion of the experimental results in section 4. Section 5 describes the photochemical model and discusses the main effects brought about by the new rate constants. The major conclusions of this work are summarized in section 6.

## 2 Experimental methods

A supersonic flow (Laval nozzle) reactor was used to perform all the experiments described here. The main features of this apparatus have been reported in earlier work.[36, 37] Modifications to the original design have allowed us to implement a detection method based on tunable narrowband radiation in the vacuum ultraviolet (VUV) wavelength range; a region where many strong electronic transitions of atomic radicals are located. In this way, it has been possible to follow the kinetics of both ground (C($^3$P),[38-40] H($^2$S)[38, 40-49] and D($^2$S), [40, 49, 50]) and excited (O($^1$D)[45, 48, 51-55] and N($^2$D)[56, 22, 23]) state atoms at low temperature. As the electronic quenching of N($^2$D) is slow with both Ar[57] and N$_2$,[14] both of these could be used as Laval nozzle carrier gases. Three nozzles were employed during the present experiments allowing four different temperatures to be attained between 50 and 177 K (one nozzle was used with both N$_2$ and Ar in separate experiments). The flow characteristics are summarized in Table 1 of Nunez-Reyes et al.[23] The apparatus was also used as a slow-flow reactor to allow measurements to be made at room temperature. N($^2$D) atoms could not be produced by photolysis during this work, due to the scarcity of appropriate precursor molecules. Instead the method described by Nunez-Reyes & Hickson[56] was employed, using the chemical reaction

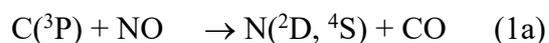
C($^3$P) + NO → N($^2$D, $^4$S) + CO   (1a)
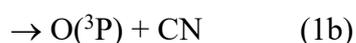
→ O($^3$P) + CN   (1b)

as a source of N($^2$D) atoms. An earlier study of the branching ratio between the two channels [N($^2$D) + N($^4$S)]/[O($^3$P)] has been estimated to be 1.5 ± 0.3 at room temperature.[58] Here, C($^3$P) atoms were produced by the 10 Hz pulsed laser photolysis of tetrabromomethane (CBr$_4$) molecules at 266 nm. The pulse energy was typically around 23 mJ with a 6 mm diameter beam. CBr$_4$ with a maximum concentration of approximately 3 × 10$^{13}$ cm$^{-3}$ was introduced into the supersonic flow by passing a small fraction of the carrier gas into a flask containing solid CBr$_4$ held at a fixed pressure and temperature. C($^1$D) is also produced by this multiphoton dissociation process with a a C($^1$D)/C($^3$P) ratio of 0.1-0.15 according to the earlier work of Shannon et al.[38] under similar conditions. During this study, N($^2$D) atoms were followed by pulsed laser induced fluorescence in the vacuum ultraviolet region (VUV LIF) through the $2s^22p^3$ $^2$D° - $2s^22p^2$($^3$P)3d $^2$F transition at 116.745 nm. Tunable light around this wavelength was obtained, initially by frequency doubling the 700.5 nm output of a pulsed tunable dye laser in a BBO crystal, generating a 350.25 nm (UV) beam with a pulse energy of approximately 8.5 mJ. After separation of the fundamental wavelength, the UV beam was focused into a cell containing 40 Torr of Xe and 560 Torr of Ar, generating tunable VUV radiation by frequency tripling. The configuration of the detection system itself (in terms of optical materials and geometries) was identical to the one used by Nunez-Reyes et al.[23] Resonant VUV LIF fluorescence from N($^2$D) atoms was collected by a solar blind photomultiplier tube (PMT), while the PMT output signal was amplified before processing by a boxcar integration system. In common with our earlier work, the first 15 microseconds following the photolysis laser pulse were not exploitable due to amplifier saturation issues. At least 70 time intervals were recorded for each temporal profile with each time point representing the average of 30 probe laser shots. Several probe laser shots were recorded prior to the photolysis laser to establish the signal baseline level. The gases used in the experiments (Linde Ar 99.999%, Xe 99.999%, $C_2H_4$ 99.95%, Air Liquide $N_2$ 99.999%, NO 99.9%) were flowed directly from the cylinders into calibrated mass-flow controllers, which were used to precisely regulate gas flows into the reactor.

## 3 Theoretical methods

To rationalize the discrepancies between the present and previous measurements and theoretical calculations, we have performed a new theoretical study of the N($^2$D) + $C_2H_4$ reaction. As the electronic state of N($^2$D) is fivefold degenerate in the absence of spin–orbit interactions in the N($^2$D) + $C_2H_4$ system, five doublet potential energy curves arise ($^2A_1$, two $^2A_2$, $^2B_1$ and $^2B_2$), which correlate with these reagents in $C_{2v}$ symmetry when N approaches the C=C bond of ethylene perpendicularly. Two $^2A'$ and three $^2A''$ states arise in Cs symmetry when N approaches the C=C bond of ethylene at other angles. As shown by Takayanagi et al.,[21] Vuitton et al.[59] and Nunez-Reyes et al.,[23] it is crucial to employ a method that allows the multiconfigurational aspect of N($^2$D) reactivity to be taken into account. Consequently, we used Complete Active Space Self-Consistent Field (CASSCF) calculations using 14 active orbitals and 15 active electrons. With the resulting molecular orbitals, the MRCI energies

(Multi-Reference Configuration Interaction) with a smaller active space (10 active orbitals and 5 active electrons) were calculated using the MOLPRO[60] suite of programs with an augmented triple zeta atomic basis set, aug-cc-VTZ. In contrast to our previous work on the N($^2$D) + $C_2H_2$ reaction,[23] the geometry was not optimized at the MCSCF level for each distance between N and the center of mass of acetylene. Instead, this was maintained at the optimized geometry when the nitrogen atom and $C_2H_4$ were separated. We compare our results with previous calculations to demonstrate the critical nature of the calculation level.

**4 Results and discussion**

Large excess concentrations of NO and $C_2H_4$ with respect to C($^3$P) (and N($^2$D) as a product of reaction 1) were used during these experiments, allowing the pseudo-first-order approximation to be applied during the data treatment and analysis. Consequently, the N($^2$D) fluorescence signal was expected to follow a biexponential temporal profile

$$I_{N(^2D)} = A(\exp(-k'_a t) - \exp(-k'_b t)) \qquad (2)$$

where A represents the theoretical maximum signal amplitude, $k'_a$ is the pseudo-first-order rate constant for N($^2$D) loss, $k'_b$ is the pseudo-first-order rate constant for N($^2$D) formation and $t$ is time. The various contributions to the constants $k'_a$ and $k'_b$ are described in Nunez-Reyes et al.,[23] replacing $C_2H_2$ by $C_2H_4$. In common with this earlier work, a single exponential function of the form

$$I_{N(^2D)} = A\exp(-k'_a t) \qquad (3)$$

was actually used to fit the fluorescence signal profiles due to the impossibility to exploit the first 15 microseconds following the photolysis laser pulse. This approximation is valid only if N($^2$D) production by reaction (1a) is essentially complete before starting the fit. In practice, this could be easily verified by plotting the logarithm of the VUV LIF signal as a function of time and by exploiting only the 'linear' part of the decay profile. Representative fluorescence profiles recorded at 296 K are shown in Figure 1.

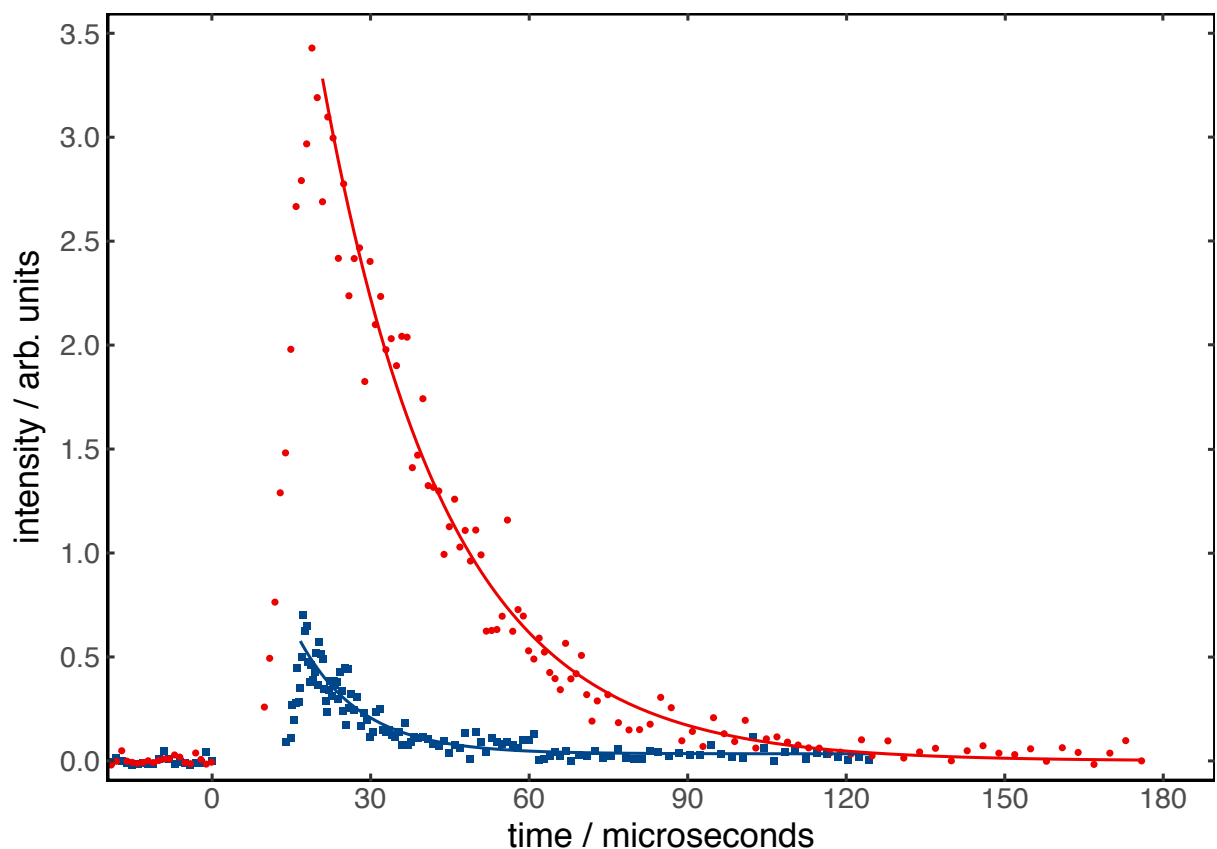

**Figure 1** $I_{N(^2D)}$ as a function of the delay between photolysis and probe lasers recorded at 127 K. (Solid red circles) without $C_2H_4$; (solid blue squares) $[C_2H_4] = 5.1 \times 10^{14}$ cm$^{-3}$. $[NO] = 4.7 \times 10^{14}$ cm$^{-3}$ for both decays. Single exponential fits to the data according to expression (3) are represented by solid red and blue lines.

**Effects of the competing C($^3$P) + C$_2$H$_4$ reaction and other secondary reactions**

Before proceeding to the next stage of the data analysis, it is first important to consider possible interferences that might arise from secondary chemistry or competing reactions within the supersonic flow. In this study, atomic carbon (in addition to N($^2$D)) is expected to react rapidly with $C_2H_4$,[61]

$$C(^3P) + C_2H_4 \rightarrow C_3H_3 + H \qquad (4)$$

in a similar manner to our earlier study of the N($^2$D) + C$_2$H$_2$ reaction,[23] where atomic carbon was also seen to react with C$_2$H$_2$. Indeed, it can be seen from Figure 1 that experiments performed with low [C$_2$H$_4$] favored the production of higher N($^2$D) atom concentrations compared to those experiments performed at high [C$_2$H$_4$]. Consequently, in order to generate enough N($^2$D) atoms to be able to accurately follow the kinetics of the N($^2$D) + C$_2$H$_4$ reaction, it was important to carefully control the C$_2$H$_4$ and NO concentrations used ([NO] is constant for any series of experiments). A more detailed discussion of this issue can be found in Nunez-Reyes et al.,[23] as well as the implications for the data analysis procedures. In addition to the effects of reaction (4) on the relative N($^2$D) signal levels, we also need to evaluate whether the products (through reaction with excess reagents NO and $C_2H_4$ in particular) might interfere with our kinetic study of the N($^2$D) + $C_2H_4$ reaction. The majority of possible secondary processes including the reaction of C($^1$D) (produced by CBr4 photolysis) with NO have been considered in earlier work and do not need to be discussed further here.[23, 56] CN radicals produced by reaction (1b) react rapidly with $C_2H_4$ leading to $C_2H_3CN$ + H as the exclusive products, although these are unlikely to affect the present kinetic measurements.[62] Another important difference to be considered, is the potential for interferences from secondary reactions of the products of reaction (4) and the equivalent C($^1$D) reaction. Propargyl radicals $C_3H_3$ and atomic hydrogen are the major products of the C($^3$P) + $C_2H_4$ reaction (0.92 ± 0.04), from experimental studies of the H-atom yield,[63] while dynamics studies of the C($^1$D) + $C_2H_4$ system indicate that $C_3H_3$ + H are also the major products of this reaction.[64]

As a resonantly stabilized species, the $C_3H_3$ radical shows only low reactivity towards stable molecules, so it does not react with $C_2H_4$. In contrast, the $C_3H_3$ + NO radical-radical reaction has been studied both experimentally[65] and theoretically.[66] The experimental investigation by DeSain et al.[65] demonstrated that the measured rate constant was highly dependent on the buffer gas pressure, consistent with a termolecular reaction to form a $C_3H_3NO$ adduct. The later theoretical study by Wang et al.[66] indicated that bimolecular product formation should be very minor, with a branching ratio of less than 6 % predicted for the HCN + $H_2CCO$ channel at a pressure of 0.1 bar. Consequently, neither of these processes are expected to interfere with our kinetic study of the N($^2$D) + $C_2H_4$ reaction. As an additional test of the experimental method, for three different temperatures (296, 177 and 127 K) decays were recorded over a range of [$C_2H_4$] at two different NO concentrations. The derived second-order rate constants were always within the experimental error bars at all three temperatures.

At any given temperature, at least 31 separate decays were recorded over a range of $C_2H_4$ concentrations. Figure 2 shows the pseudo-first-order rate constant (extracted using fits such as those shown in Figure 1) plotted as a function of [$C_2H_4$], allowing the second-order rate constant to be derived from the slopes of weighted linear-least squares fits to the data.

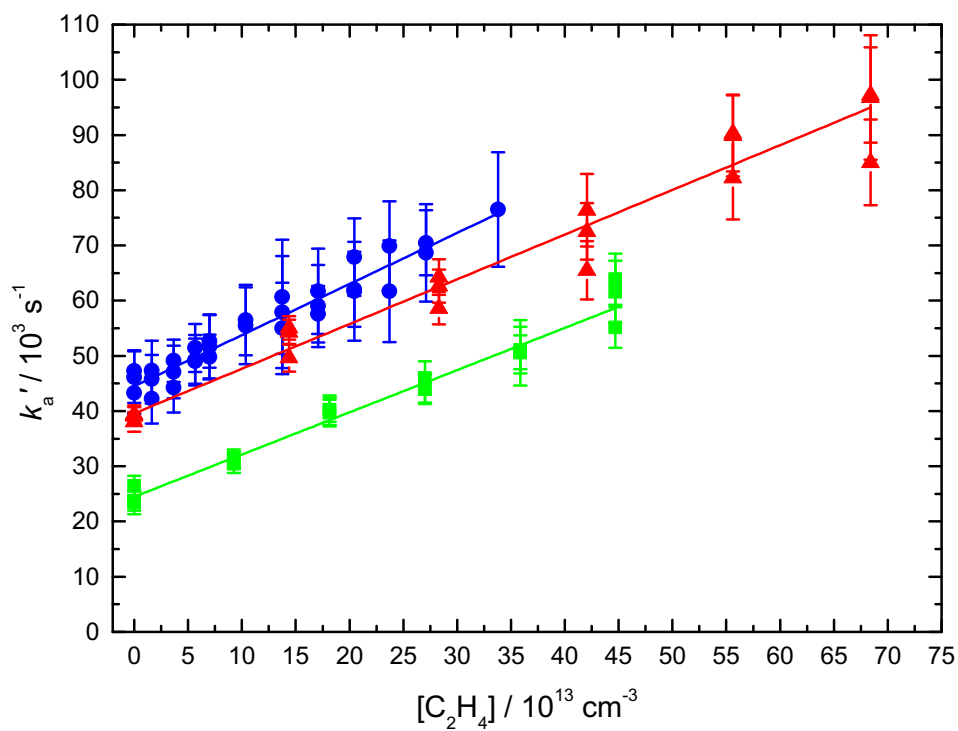

**Figure 2** Derived pseudo-first-order rate constant $k_a'$ as a function of [$C_2H_4$]. (Red solid triangles) 296 K; (green solid squares) 177 K; (blue solid circles) 50 K. Solid lines represent weighted linear least-squares fits to the data. The errors on individual data points are shown at the level of a single standard deviation and are determined from fits to temporal decays similar to those displayed in Figure 1 using expression (3).

The derived second-order rate constants and the results of earlier studies are displayed as a function of temperature in Figure 3. The values obtained during this work are also listed in Table 1 alongside other related information.

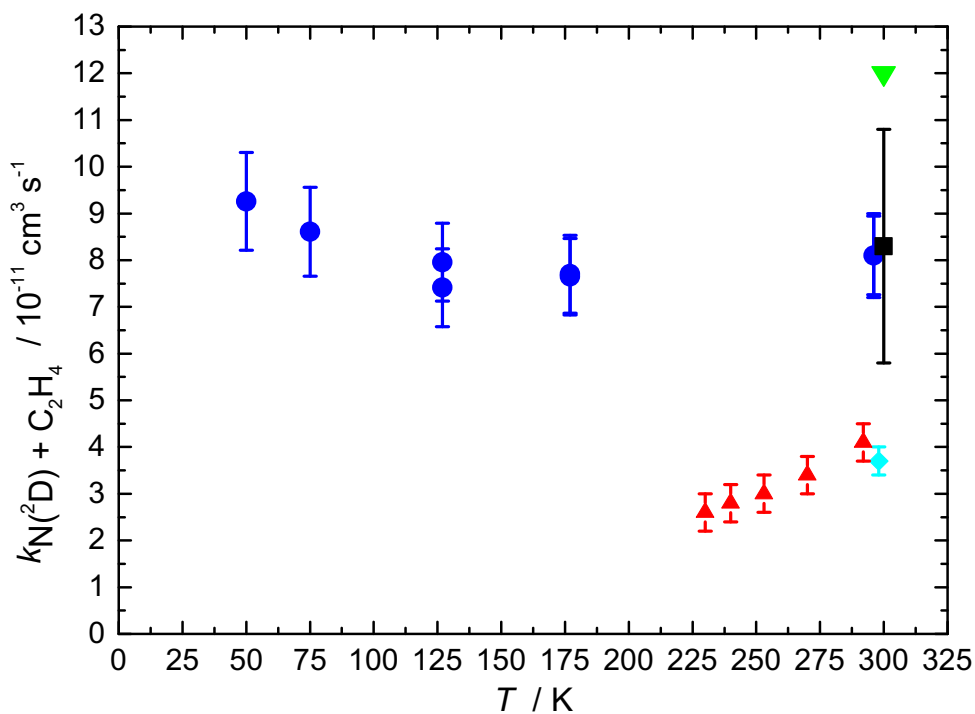

**Figure 3** Second-order rate constants for the N($^2$D) + C$_2$H$_4$ reaction as a function of temperature. (Green triangle) Black et al.;[15] (black square) Fell et al.;[16] (cyan diamond) Sugawara et al.;[20] (red triangles) Sato et al.;[24] (blue circles) this work. Error bars on the present values represent the statistical (1σ) and systematic uncertainties (estimated to be 10 %).

**Table 1** Measured rate constants for the N($^2$D) + C$_2$H$_4$ reaction

| T / K | N[b] | [C$_2$H$_4$] / 10$^{14}$ cm$^{-3}$ | [NO] / 10$^{14}$ cm$^{-3}$ | $k_{N(^2D)+C_2H_4}$ / 10$^{-11}$ cm$^3$ s$^{-1}$ |
|---|---|---|---|---|
| 296 | 18 | 0 – 6.8 | 6.4 | (8.1 ± 0.8)[c] |
| 296 | 18 | 0 – 6.8 | 4.3 | (8.1 ± 0.9) |
| 177 ± 2[a] | 18 | 0 – 4.5 | 2.8 | (7.7 ± 0.8) |
| 177 ± 2 | 18 | 0 – 4.5 | 4.2 | (7.6 ± 0.8) |
| 127 ± 2 | 18 | 0 – 5.1 | 3.2 | (7.4 ± 0.8) |
| 127 ± 2 | 18 | 0 – 5.1 | 4.7 | (8.0 ± 0.8) |
| 75 ± 2 | 36 | 0 – 2.4 | 2.7 | (8.6 ± 1.0) |
| 50 ± 1 | 31 | 0 – 3.4 | 4.2 | (9.3 ± 1.0) |

[a]Uncertainties on the calculated temperatures represent the statistical (1σ) errors obtained from Pitot tube measurements of the impact pressure. [b]Number of individual measurements. [c]Uncertainties on the measured rate constants represent the combined statistical (1σ) and estimated systematic (10%) errors.

There are three earlier measurements of the rate constant for the N($^2$D) + C$_2$H$_4$ reaction at room temperature. Black et al.[15] used the flash photolysis of N$_2$O at 147 nm as a source of N($^2$D) atoms in their investigation of N($^2$D) deactivation with several collision partners. The kinetics of N($^2$D) loss was followed by detecting the emission from NO in the B$^2\Pi$ electronic state, where NO(B$^2\Pi$) was produced through the N($^2$D) + N$_2$O reaction. They determined a rate constant $k_{N(^2D)+C_2H_4}$(300 K) = (1.2 ± 0.3) × 10$^{-10}$ cm$^3$ s$^{-1}$. Sugawara et al.[20] employed the pulsed radiolysis technique to form N($^2$D) from N$_2$ coupled with resonance absorption detection of N($^2$D) through its atomic transitions around 149.3 nm (the three fine structure lines of the $^2$D → $^2$P$_J$ transition could not be resolved in this work). They measured a rate constant of (3.7 ± 0.3) × 10$^{-11}$ cm$^3$ s$^{-1}$ for the reaction of N($^2$D) with C$_2$H$_4$ at 298 K. In their fast-flow reactor study, Fell et al.[16] produced N($^2$D) atoms using the microwave discharge of N$_2$ in helium carrier gas. These atoms were followed directly by electron spin resonance spectroscopy. They determined rate constants for numerous deactivation processes involving N($^2$D), including the N($^2$D) + C$_2$H$_4$ reaction, obtaining a rate constant of (8.3 ± 2.5) × 10$^{-11}$ cm$^3$ s$^{-1}$ at 298 K; a value that is in excellent agreement with the values measured during this study at 296 K of (8.1 ± 0.8) × 10$^{-11}$ cm$^3$ s$^{-1}$ and (8.1 ± 0.9) × 10$^{-11}$ cm$^3$ s$^{-1}$. More recently, Sato et al.[24] studied the kinetics of this reaction over the 230-292 K range using essentially the same technique as the one used by Sugawara et al.[20] They determined a rate constant of (4.1 ± 0.4) × 10$^{-11}$ cm$^3$ s$^{-1}$ at 292 K in good agreement with the previous work of Sugawara et al.[20] The reactivity was seen to decrease at lower temperature, with the rate constant reaching a value of (2.6 ± 0.4) × 10$^{-11}$ cm$^3$ s$^{-1}$ at 230 K. Based on the measured rate constants, Sato et al.[24] derived the Arrhenius parameters, the preexponential factor, A = (2.3 ± 0.3) × 10$^{-10}$ cm$^3$ s$^{-1}$ and the activation energy, $E_a$ = 4.184 ± 0.4 kJ mol$^{-1}$. In order to provide a quantitative description of the temperature dependence of the N($^2$D) + C$_2$H$_4$ reaction, Sato et al.[24] also performed canonical variational transition state theory (VTST) calculations, based on ab initio calculations performed initially at the CASSCF(5,5)/cc-pVDZ level of theory. These calculations resulted in classical barrier height of 13.4 kJ mol$^{-1}$, much larger than the experimental activation energy. Instead, these authors used a more accurate technique, namely the second-order configuration interaction (SOCI) method to derive the energies along the intrinsic reaction coordinate path, leading to a lower barrier height of 2.9 kJ mol$^{-1}$ calculated at the SOCI/cc-pVDZ level of theory. Using the SOCI results, the VTST calculations of Sato et al.[24] were seen to reproduce the experimental results only if all five of the doublet electronic surfaces that correlate with N($^2$D) + C$_2$H$_4$ reagents contribute to the overall reactivity, as given by an electronic partition function of 1.0 (instead of the expected value of 0.2 for a single surface). Sato et al.[24] justified this assumption by suggesting that nonadiabatic transitions could be important in the asymptotic region where shallow van der Waals wells were calculated to exist for all five PESs.

The previous calculations by Sato et al.[24] clearly show the importance of the choice of the method used to calculate the potential energy along the reaction coordinate for N($^2$D) attack on ethylene. Our present ab-initio calculations at the CASSCF level (14 active orbitals and 15 active electrons) + MRCI (10 active orbitals and 5 active electrons) with an augmented VTZ basis show no barrier for the $^2B_1$ and $^2B_2$ states in $C_{2v}$ symmetry (N approach perpendicular to the C=C bond of ethylene) while the other three electronic states are strongly repulsive (see Figure 4). In our calculations the distances and angles of $C_2H_4$ were fixed at the $C_2H_4$ equilibrium geometry. Consequently, the potential surfaces are not the minimum ones except at long distance, although this does not change the absence of a barrier for the $^2B_1$ and $^2B_2$ curves.

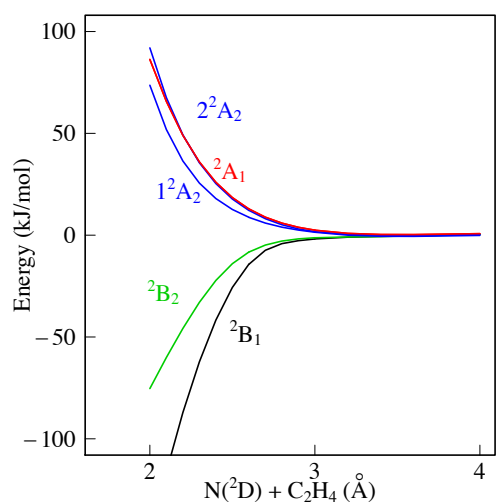

**Figure 4** Profiles of the five electronic state potential curves for the N($^2$D) + $C_2H_4$ reaction calculated at the MRCI/AVTZ level of accuracy.

As already shown for the N($^2$D) + $C_2H_2$ reaction,[23] a correct description of N($^2$D) reactions requires the use of evolved theoretical methods that allow for the multiconfigurational aspect. Our results clearly show that there is little doubt that the N($^2$D) + $C_2H_4$ reaction is barrierless for two of the five doublet potential energy curves leading to a partition function of 0.4. The difference between our calculations and the ones of Sato et al.,[24] who found a small barrier, shows the importance of using a complete active space in CASSCF calculations associated with a multi-reference configuration interaction method. Future calculations of the rate constant would require a complete description of the first two PESs while also including non-adiabatic effects at long distance. Nevertheless, the large measured rate constant at low temperature is clearly indicative of a barrierless reaction. Further statistical calculations should be performed on this system, by employing these new energies to examine the effect on the calculated rate constants.

Using the Arrhenius parameters provided by Sato et al.[24] and extrapolating to lower temperature leads to values of $k_{N(^2D)+C_2H_4}$(177 K) and $k_{N(^2D)+C_2H_4}$(127 K) of $1.3 \times 10^{-11}$ cm$^3$ s$^{-1}$ and $4.4 \times 10^{-12}$ cm$^3$ s$^{-1}$ respectively. This can be compared with average values measured here of $k_{N(^2D)+C_2H_4}$(177 K) = $(7.7 \pm 0.8) \times 10^{-11}$ cm$^3$ s$^{-1}$ and $k_{N(^2D)+C_2H_4}$(127 K) = $(7.7 \pm 0.8) \times 10^{-11}$ cm$^3$ s$^{-1}$; respectively six and eighteen times larger than those derived from the Sato et al.[24] measurements. As the temperature dependence of the rate constant measured in the present work is fairly weak, we prefer to recommend a temperature independent value of $8.1 \times 10^{-11}$ cm$^3$ s$^{-1}$. Assuming an average temperature of 170 K for the atmosphere of Titan leads to a rate constant that is almost seven times larger than the currently recommended value of Dutuit et al.[10] It should be noted that a similar temperature dependence to the one measured by Sato et al.[24] for the N($^2$D) + C$_2$H$_4$ reaction was also derived by this group for the N($^2$D) + C$_2$H$_2$ reaction,[21] in contrast with our recent experimental and theoretical study of this reaction.[23] Consequently, the present work casts further doubt on the results of these earlier measurements that seem to be incompatible with the predicted barrierless nature of these reactions.

## 5 Photochemical model

In order to test the effect of these measurements on the chemistry of Titan's atmosphere, we included the new rate constants for the N($^2$D) + C$_2$H$_4$ reaction in a photochemical model. The 1D-model of Dobrijevic et al.[67] was used as the basis for this work, which treats the coupled chemistry of neutrals and cations from the lower atmosphere to the ionosphere. This model also includes an updated chemistry of aromatic compounds,[68] while earlier work by these authors had already focused on the neutral chemistry of nitrogen bearing molecules.[31, 33] Recent updates to the reactions of N($^2$D) with saturated and unsaturated hydrocarbons are treated elsewhere.[22, 23] Here, we have also introduced new product channels for the N($^2$D) + C$_2$H$_4$ reaction considering the recent theoretical and experimental work of Balucani et al.[27] leading mainly to the formation of CH$_2$NCH and c-CH$_2$NCH. These species were introduced into the chemical network as well as their most important loss processes (including photodissociations calculated at the EOM-CCSD(T) level (see the electronic supplementary information)). To calculate the rate constant values for these new reactions, DFT calculations were employed with the M06-2X functional and the AVTZ basis using the Gaussian suite program.[69] To build the network we first considered the reactions of $CH_2NCH$ and c-$CH_2NCH$ with hydrogen atoms and $CH_3$ which are the most abundant reactive species in Titan's atmosphere. $CH_2NCH$ is predicted to react quickly with H as there is no (or a very small) barrier at the M06-2X/AVTZ level leading only to HCN + $CH_3$ with very little production of $CH_3CN$ isomers. As $CH_2NCH$ reacts quickly with hydrogen atoms there is no need to consider its reactions with other minor reactive species such as $C_2H$ and CN. The reaction of $CH_2NCH$ with $CH_3$ shows a relatively large barrier equal to 28 kJ/mol at the M06-2X/AVTZ level and is neglected here. The cyclic isomer, c-$CH_2NCH$, shows a significantly lower reactivity towards hydrogen atoms than the linear isomer, with a barrier of approximately 18 kJ/mol. As c-$CH_2NCH$ is expected to react relatively slowly with atomic hydrogen, associated with relatively inefficient photodissociation, we introduced the reactions of c-$CH_2NCH$ with $N(^2D)$, $C_2H$ and CN. For all these reactions, we considered addition to the C=N double bond and neglected the abstraction pathways. With the new $N(^2D)$ + $C_2H_4$ product branching ratios, and the subsequent reactions of $CH_2NCH$ isomers, HCN and $CH_3$ are the favored products, in contrast to the results of previous simulations where $CH_3CN$ was the major product. To illustrate the effect of the rate change we have performed two simulations by varying some rate constants. The first set of simulations performed in this study adopted the previously recommended rate constant for the $N(^2D)$ + $C_2H_4$ reaction (where $\alpha = 2.35 \times 10^{-10}$, $\beta = 0$ and $\gamma = 503$ in the modified Arrhenius expression $k(T) = \alpha(T/300)^\beta e^{-\gamma/T}$). The second set of simulations, which corresponds to our nominal model, replaced the rate constant values by a temperature independent value of $8.1 \times 10^{-11}$ cm$^3$ s$^{-1}$ (see Table S1). The differences produced by the two models for various nitrogen-bearing species are summarized in Table 2.

**Table 2** Mole fractions given by the new model and ratios given by the new model / old model (limited to species with mole fractions greater than $1.0 \times 10^{-12}$ in the upper atmosphere).

| Altitude | 300 km | | 600 km | | 800 km | | 1000 km | | 1200 km | |
|---|---|---|---|---|---|---|---|---|---|---|
| Species | Mole fraction | $R^b$ | Mole fraction | $R$ | Mole fraction | $R$ | Mole fraction | $R$ | Mole fraction | $R$ |
| $CH_3CN$ | 2.3(-8)[a] | 2.9 | 1.2(-6) | 3.1 | 7.4(-7) | 3.1 | 1.7(-6) | 3.0 | 1.4(-6) | 2.9 |
| c-$CH_2NCH$ | 3.3(-11) | 6.0 | 7.2(-9) | 6.0 | 2.4(-7) | 6.1 | 2.4(-6) | 5.9 | 4.3(-6) | 6.2 |
| $C_3H_5CN$ | 1.5(-10) | 2.5 | 1.8(-10) | 2.6 | 3.4(-10) | 2.8 | 4.4(-10) | 3.4 | 5.5(-11) | 3.0 |
| $CH_2CN$ | 2.2(-14) | 2.4 | 1.9(-12) | 2.9 | 1.5(-8) | 2.6 | 3.4(-6) | 3.3 | 5.8(-6) | 3.3 |
| $C_2H_5CN$ | 8.8(-9) | 1.7 | 3.6(-8) | 2.5 | 2.2(-7) | 3.1 | 1.5(-7) | 3.1 | 1.3(-8) | 3.2 |
| $C_3H_7CN$ | | | 5.0(-11) | 1.5 | 4.6(-10) | 3.1 | 7.8(-10) | 3.5 | 4.6(-11) | 3.5 |
| $CH_2NCH$ | | | | | 3.1(-13) | 6.7 | 1.6(-8) | 5.9 | 1.1(-6) | 6.6 |
| $C_2H_4CN$ | | | | | 2.9(-13) | 2.4 | 2.0(-10) | 3.3 | 3.0(-10) | 3.2 |
| $HC_2N_2$ | | | | | | | 2.0(-9) | 2.8 | 4.8(-8) | 3.0 |
| CH$_3$CNH$^+$ | | | | | | 5.8(-12) | 2.3 | 5.0(-9) | 3.3 | 2.6(-8) | 3.3 |
| C$_2$H$_5$CNH$^+$ | | | | | | 1.1(-11) | 3.6 | 1.0(-10) | 3.3 | 1.8(-11) | 3.2 |
| CH$_2$CNH$^+$ | | | | | | | | 2.4(-9) | 3.1 | 7.9(-9) | 3.3 |

[a]2.0(-8) ≡ 2.0 × 10$^{-8}$. [b]Ratio of new mole fraction/old mole fraction.

The new rate constants induce notable increases for both c-CH$_2$NCH and CH$_3$CN. This result was expected for c-CH$_2$NCH as the N($^2$D) + C$_2$H$_4$ reaction is the only source of c-CH$_2$NCH, and thus an increase in the rate of this reaction leads to an increase in its production. The increase in CH$_3$CN is less obvious to understand because the new branching ratios of the N($^2$D) + C$_2$H$_4$ reaction do not lead directly to CH$_3$CN. However, the N($^2$D) + C$_2$H$_4$ reaction is still a non-negligible source of CH$_3$CN, via the H + c-CH$_2$NCH reaction, because the new rate constant for the N($^2$D) + C$_2$H$_4$ reaction is much larger at low temperature than the previously recommended values based on the measurements of Sato et al.[24] Indeed, the N($^2$D) + C$_2$H$_4$ reaction becomes the main consumption pathway of N($^2$D), followed by the N($^2$D) + C$_2$H$_2$, N($^2$D) + CH$_4$ and N($^2$D) + HCN reactions, while the flux of c-CH$_2$NCH production (and consequently CH$_3$CN production) are important. The increase in the CH$_3$CN abundance leads to an increase in CH$_2$CN, the main photodissociation product of CH$_3$CN, followed by an increase of chemically related species such as C$_2$H$_5$CN (which is produced through the CH$_3$ + CH$_2$CN reaction). The calculated CH$_3$CN, C$_2$H$_5$CN and c-CH$_2$NCH abundances in Titan's atmosphere with the updated chemical network are presented in Figure 5 alongside available observations.

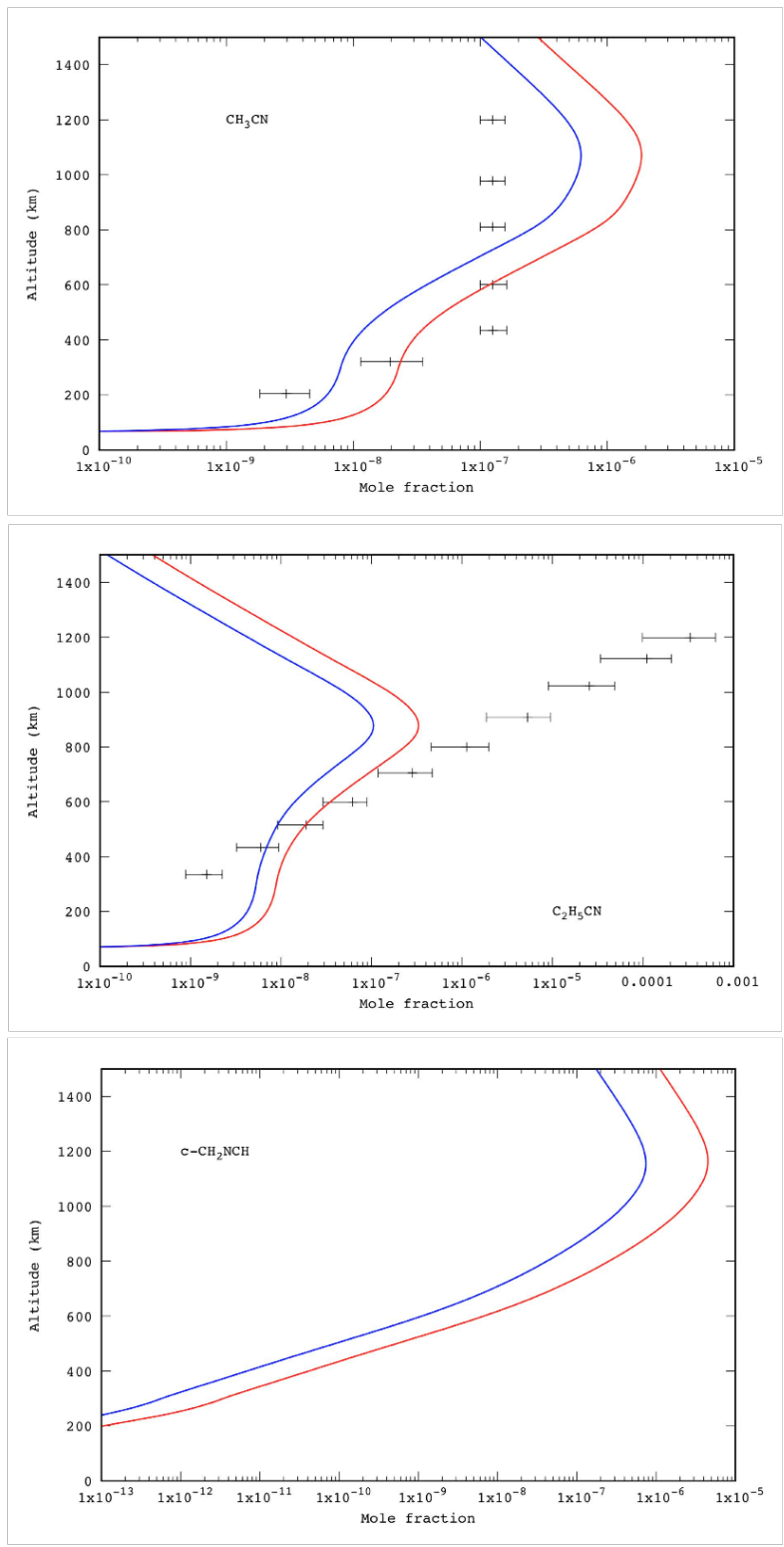

**Figure 5** Mole fractions of $CH_3CN$, $C_2H_5CN$ and $c-CH_2NCH$ as a function of altitude with the old rate constant (blue curves) and new rate constant (red curves). The observations are from Cordiner et al 2019.[70]

Blue curves represent the simulations using the currently recommended rate constants (but considering the new products of the N($^2$D) + $C_2H_4$ reaction and the subsequent reactions of these products) while the red curves represent the simulations employing the new temperature independent rate constant.

The new rate for the N($^2$D) + $C_2H_4$ reaction significantly modifies the altitudinal profile for $CH_3CN$. Comparison with the analysis of Cordiner et al. 2019[70] is not straightforward.

In particular, the constant profile derived from ALMA observations is not in agreement with our model. This might be due to the fact that ALMA cannot efficiently constrain the mole fraction profile of $CH_3CN$ above an altitude of 500-600 km. We are currently developing a radiative transfer model at submillimetre wavelengths adapted to Titan's atmosphere[71] that will allow us to test our profile by directly comparing synthetic spectra to ALMA spectra, although this work is beyond the scope of the present paper.

For $C_2H_5CN$ the agreement is slightly improved but two phenomena are still not well described; the abundance in the upper atmosphere is largely underestimated and the shape of the curve in the lower atmosphere is not well described as the simulations predict a peak around an altitude of 800 km in the $C_2H_5CN$ abundance which is not present in the observations. The underestimation of $C_2H_5CN$ in the upper atmosphere may be due to the lack of radiative association reactions in our model. The theoretical flat profile of $C_2H_5CN$ in the lower atmosphere, similar to HCN and $CH_3CN$ profiles, is very likely due to the lack of efficient consumption reactions of $C_2H_5CN$. These missing reactions could include the depletion of $C_2H_5CN$ on aerosols and/or its reaction with atomic hydrogen, which would need to be studied experimentally and theoretically.

Despite the fact that these simulations show that c-$CH_2NCH$ is produced in significant quantities by the N($^2$D) + $C_2H_4$ reaction, its concentration probably does not reach a high enough level in the lower atmosphere (due to its moderate reactivity with hydrogen atoms) to be detected by infrared spectroscopy. However, as the electric dipole moment of c-$CH_2NCH$ is calculated to be fairly strong (2.2 Debye at the M06-2X/AVTZ level) it would be interesting to try to detect c-$CH_2NCH$ through rotational spectroscopy such as with the ALMA interferometer (Atacama Large Millimeter/Submillimeter Array). It should be noted, however, that the uncertainties on the chemistry of c-$CH_2NCH$ are very large, both for the production route (the branching ratio of the N($^2$D) + $C_2H_4$ reaction towards c-$CH_2NCH$ is not well determined) and for its reactivity with atomic hydrogen.

## 6 Conclusions

This work reports the result of a joint experimental and theoretical investigation of the N($^2$D) + C$_2$H$_4$ reaction. On the experimental side, a supersonic flow reactor was used to measure rate constants for this process over the 50-296 K temperature range. N($^2$D) atoms were generated as a product of the C + NO reaction during this work, with C-atoms themselves being produced by the 266 nm pulsed laser photolysis of CBr$_4$. N($^2$D) atoms were detected directly by pulsed laser induced fluorescence at 116.7 nm. The reaction is seen to occur rapidly down to 50 K, displaying little or no temperature dependence, in contrast with earlier work above 230 K. On the theoretical side, electronic structure calculations employing a multiconfigurational method with a large active space, showed that two of the five doublet surfaces correlating with reagents present no barriers to product formation. As with the experimental work, the new calculations disagree with earlier theoretical studies of the N($^2$D) + C$_2$H$_4$ reaction at a lower level that derived a barrier for this process. Such calculations as applied to N($^2$D) reactivity are clearly highly method dependent. The new rate constants were introduced into a photochemical model of Titan's atmosphere to test its effect on atmospheric species. As the previously recommended products of this reaction have since been shown to be incorrect, we also included the updated product channels in these simulations alongside the most important loss processes for the products themselves. After the update, the N($^2$D) + C$_2$H$_4$ reaction becomes the dominant loss process for N($^2$D) atoms in Titan's atmosphere. Furthermore, several nitrogen bearing species including nitrile compounds CH$_3$CN, C$_2$H$_5$CN, C$_3$H$_7$CN and C$_3$H$_5$CN display large increases of their predicted mole fractions, highlighting the importance of the N($^2$D) + C$_2$H$_4$ reaction to the overall photochemistry.

## Conflicts of interest

There are no conflicts to declare.

## Acknowledgements

K. M. H. acknowledges support from the French program ''Physique et Chimie du Milieu Interstellaire'' (PCMI) of the CNRS/INSU with the INC/INP co-funded by the CEA and CNES as well as funding from the ''Program National de Planétologie'' (PNP) of the CNRS/INSU. C. B. acknowledges support from the project OSCAR ANR-15-CE29-0017 of the French National Research Agency (ANR) through the provision of a postdoctoral fellowship.